# On-chip integrated waveguide amplifiers on Erbium-doped thin film lithium niobate on insulator

**Junxia Zhou,**[1,2] **Youting Liang,**[1,2] **Zhaoxiang Liu,**[2] **Wei Chu,**[2] **Haisu Zhang,**[2] **Difeng Yin,**[3,4] **Zhiwei Fang,**[2,8] **Rongbo Wu,**[3,4] **Jianhao Zhang,**[3,4] **Wei Chen,**[2] **Zhe Wang,**[3,4,5] **Yuan Zhou,**[3,4] **Min Wang,**[2] **and Ya Cheng**[1,2,6,7,9]

[1]*State Key Laboratory of Precision Spectroscopy, East China Normal University, Shanghai 200062, China*
[2]*XXL—The Extreme Optoelectromechanics Laboratory, School of Physics and Electronic Science, East China Normal University, Shanghai 200241, China*
[3]*State Key Laboratory of High Field Laser Physics and CAS Center for Excellence in Ultra-intense Laser Science, Shanghai Institute of Optics and Fine Mechanics (SIOM), Chinese Academy of Sciences (CAS), Shanghai 201800, China*
[4]*Center of Materials Science and Optoelectronics Engineering, University of Chinese Academy of Sciences, Beijing 100049, China*
[5]*School of Physical Science and Technology, ShanghaiTech University, Shanghai 200031, China*
[6]*Collaborative Innovation Center of Extreme Optics, Shanxi University, Taiyuan 030006, China.*
[7]*Collaborative Innovation Center of Light Manipulations and Applications, Shandong Normal University, Jinan 250358, People's Republic of China*
[8]*zwfang@phy.ecnu.edu.cn*
[9]*ya.cheng@siom.ac.cn*

We demonstrate on-chip light amplification with integrated optical waveguide fabricated on erbium-doped thin film lithium niobate on insulator (TFLNOI) using the photolithography assisted chemo-mechanical etching (PLACE) technique. A maximum internal net gain of 18 dB in the small-signal-gain regime is measured at the peak emission wavelength of 1530 nm for a waveguide length of 3.6 cm, indicating a differential gain per unit length of 5 dB/cm. This work paves the way to the monolithic integration of diverse active and passive photonic components on the TFLNOI platform.

http://dx.doi.org/XX.XXXX

## 1. INTRODUCTION

The establishment of erbium-doped fiber amplifier (EDFA), featuring a low-nonlinearity and low-noise amplification with a broad gain covering the telecom C- and L-bands, underpins the current long-haul fiber lightwave systems. The essential advantages of rare-earth-ion-doped materials at longer excited-state lifetimes and less refractive index changes induced by excitations of doped ions compared to the electron-hole pairs in III-V semiconductors, also boost deep and extensive researches on the photonic integration of erbium-doped waveguide amplifiers and lasers with a variety of passive components on a single chip [1, 2]. An abundant of host materials including crystalline materials like lithium niobate and amorphous materials like aluminum oxides, are widely investigated to provide spatially and temporally stable gain by erbium doping for high-bit-rate optical communication and narrow-linewidth laser source [2-7].

Recently, the thin film lithium niobate on insulator (TFLNOI) platform has attracted increasing interests in photonic integrated circuits (PIC) owing to its broad optical transparency window (0.35-5 μm), high nonlinear coefficient ($d_{33} = -41.7 \pm 7.8\ pm/V@\lambda = 1.058$ μm), high refractive index (~2.2), and large electro-optical effect ($r_{33} = 30.9\ pm/V@\lambda = 632.8$ nm). Significant progresses in the design and fabrication of passive photonic components based on TFLNOI have been made towards both scientific and technical applications including nonlinear frequency conversion, high-speed integrated modulator, quantum information technology, microwave photonics, frequency microcomb and precision metrology [8-14]. A growing impetus on integration of active components such as lasers and amplifiers based on the erbium-doped TFLNOI platform emerges, though the microchip lasers on the erbium-doped TFLNOI are only demonstrated very recently, showing great potentials for high-performance scalable light sources on the TFLNOI platform [15, 16].

Here, we demonstrate the first monolithically integrated erbium-doped TFLNOI waveguide amplifier, to the best of our knowledge, fabricated by photolithography assisted chemo-mechanical etching (PLACE). The on-chip integrated device exhibits more than 40 nm bandwidth of internal net gain around the C-band, with the peak gain of 18 dB in the small-signal gain regime at ~1530 nm for a 3.6 cm long waveguide. This work demonstrates the efficient broadband amplifications by erbium-doped TFLNOI waveguide amplifiers, paving the way to the monolithic integration of diverse active and passive photonic components on the TFLNOI platform.

## 2. EXPERIMENTAL RESULTS

Fig. 1(a) shows the schematic of the monolithic $Er^{3+}$-doped lithium niobate (LN) waveguide amplifier chip. The 600-nm-thick $Er^{3+}$-doped Z-cut LN waveguides with a top-width of ~1.2 μm and a bottom-width of ~4 μm are located on the top of a 2-μm-thick silica layer, which is further bonded onto a 0.5-mm-thick undoped crystalline LN wafer. The monolithic $Er^{3+}$-doped LN waveguide amplifier chip was fabricated using PLACE technique, and details about the fabrication process can be found in Refs. 17-20. The concentration of $Er^{3+}$ ions in the LN waveguides is 1 mol%. Fig. 1(b) and (c) presents the zoom-in images of the curved and straight waveguides of the on-chip $Er^{3+}$-doped LN waveguide amplifier, respectively. The rim of the LN waveguide displays interference patterns under illumination, indicating the varying thickness at the edge of the LN waveguide. The spiral design reduces the overall footprint of the amplifier, providing a total gain length of 3.6 cm with a minimum bending radius of 800 μm. Fig. 1 (d) presents the scanning electron microscope (SEM) image of the cross section of the fabricated $Er^{3+}$-doped LN waveguide, the tilt angle of SEM image is about 52°. The LN waveguide is coated with a thin layer of metal film for the sake of clear imaging in the SEM process, thus both the top surface and sidewall appear a little bit rough. Fig.1 (e) shows the simulated electric field distribution of the fundamental mode in the LN waveguide at λ=1550nm.

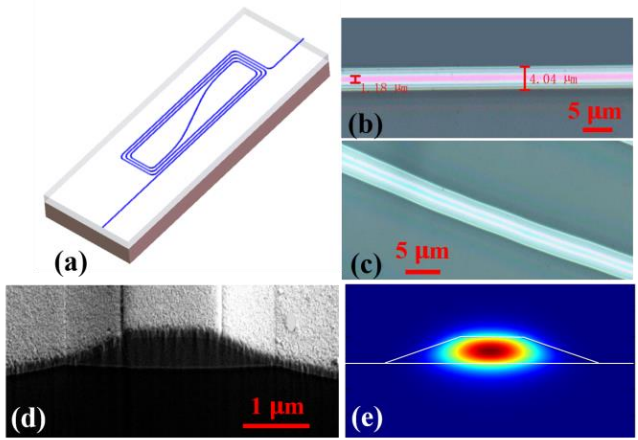

Fig. 1. (a) 3D schematic of the $Er^{3+}$-doped LN waveguide amplifier chip. Optical micrograph of the straight (b) and (c) curved waveguides. (d)SEM images of the cross sections of the fabricated $Er^{3+}$-doped LN waveguide (tilt angle: 52°). (e) The simulated electric field distribution of single mode in the LN waveguide at λ=1550nm.

The pump-and-signal method was applied to measure the internal net gain in the $Er^{3+}$-doped LN waveguides as shown in Fig. 2. The pump light at 980 nm is provided by a diode laser (CM97-1000-76PM, Wuhan Freelink Opto-electronics Co., Ltd.), while a continuous-wave C-band tunable laser (TLB 6728, New Focus Inc.) with the wavelength range from 1520 nm to 1570 nm was used as the signal. The polarization states of both the pump and signal lasers are adjusted using several in-line fiber polarization controllers. The pump and signal light waves were combined (seperated) by the fiber-based wavelength division multiplexers (WDM) at the input (output) port of the integrated amplifier. Bidirectional pumping scheme was employed to invert the erbium ions more uniformly along the full length of the amplifier. The polarization states of the pump and signal lasers were adjusted by an commercialized fiber polarization controller. Both the output amplified signals and amplified spontaneous emissions (ASE) were measured by an optical spectrum analyzer (OSA: AQ6370D, YOKOGAWA Inc.). Since the launching and collecting lensed fibers have exactly the same characteristics in all the experiments, indentical coupling efficiencies were assumed for both the input and output ports. The photograph of a pumped ($λ_P$ = 980 nm) $Er^{3+}$-doped LN waveguide spiral amplifier chip in Fig. 2 displays the strong green upconversion fluorescence along the $Er^{3+}$-doped LN waveguide.

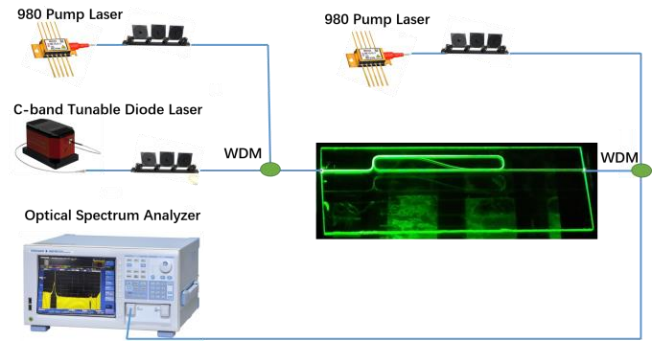

Fig. 2. Experimental setup for the gain measurements on the $Er^{3+}$-doped LN waveguide spiral amplifier. The photograph of the TFLNOI chip shows the characteristic green light emission of erbium when pumping the $Er^{3+}$-doped LN waveguide.

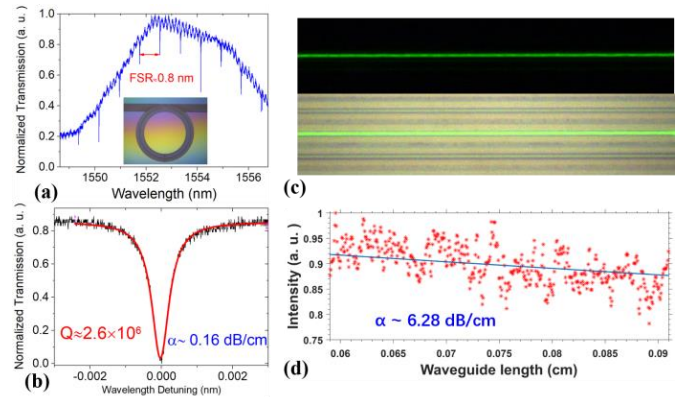

Fig. 3 Propagation losses measurement, (a) Transmission spectrum of the $Er^{3+}$-doped LN microring, the inset displays a 400-μm-diameter $Er^{3+}$-doped LN microring integrated with a undoped LN waveguide. (b)The Lorentzian fitting indicating the Q-factors of 2.6×10$^6$ of the microring as measured at 1554.13 nm wavelength. (c) Top-view microscope Image of a 300-μm-long straight waveguide. (d)Intensity distribution along the propagation direction in the 300-μm-long straight waveguide.

The optical propagation losses of the $Er^{3+}$-doped LN waveguides around the signal wavelength (~1550 nm) were first characterized using whispering-gallery-resonator-loss measurements [17]. An $Er^{3+}$-doped LN microring resonator with the diameter of 400 μm was fabricated using the same waveguide parameters shown in Fig. 1. The measured transmission spectrum of the micro-ring resonator around 1550 nm is depicted in Fig. 3(a), featuring a free spectral range (FSR) of 0.8 nm which is well consistent with the 400 μm diameter of the microring resonator. The regularly spaced resonance lines indicate the dominant excitations of the fundamental modes in the microring resonator. As shown in Fig.3(b), one of the whispering-gallery modes at the resonance wavelength of 1554.13 nm was chosen for the measurement of the loaded Q-factor by fitting the transmission curve with a Lorentz function, giving $Q \approx 2.6 \times 10^6$. Meanwhile, the group index at the wavelength of 1554.13 nm was calculated by group index $n_g = \lambda^2/(\pi D \cdot FSR)$ to be 2.4, where D is the diameter of the microring resonator. Consequently, the propagation loss of the $Er^{3+}$-doped LN waveguide was deduced to be 0.16 dB/cm using the expression $\alpha_l = 2\pi n_g/(Q\lambda)$. It should be noted that the retrieved propagation loss includes both the waveguide scattering loss due to the sidewall roughness and the absorption loss induced by ground-state erbium ions.

The optical losses at the pump wavelength of 980 nm were further estimated by measuring the green fluorescence from the energy-transfer-upconversion (ETU) of $Er^{3+}$ ions excited by the pump light [2]. Top-view microscope images of the scattered fluorescence (~530nm) in a 300-μm-long straight waveguide under the dark-field and bright-field conditions captured by an COMS camera (DCC3240C, Thorlabs Inc.) were shown in Fig. 3(c). Assuming that in average the intensity of scattered fluorescence is proportional to the local intensity of the pump travelling through the waveguide, a Lambert-Beer law for the decay of the fluorescence along the waveguide path is observed and fitted as plotted in Fig. 3(d), indicating a propagation loss of 6.28 dB/cm around 980 nm. Besides, the fiber-to-chip coupling losses for the integrated amplifier were measured to be 10 dB and 7.3 dB per facet at the wavelengths of 980 nm and 1550 nm, respectively. The high coupling losses are due to the unoptimized mode field profiles in both of the LN waveguide and lensed fibers.

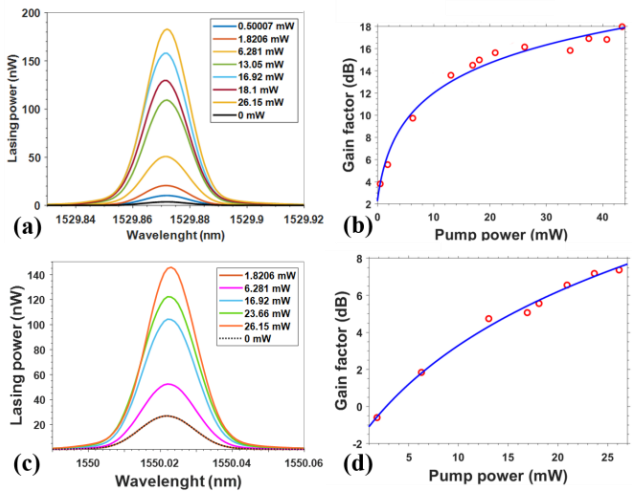

Fig. 4 Gain characterization of the $Er^{3+}$-doped LN waveguides. Measured spectra from the OSA at the wavelength of 1530 nm (a) and 1550 nm (c) at different pump powers. Net gains of the $Er^{3+}$-doped LN waveguides as a function of launched pump power for signal wavelengths of 1530 nm (b) and 1550nm (d). The blue solid lines in (b, d) are guided for the eyes.

The internal net gain of the $Er^{3+}$-doped LN waveguide amplifier was measured by the signal enhancement method and defined by the following equation

$$g = 10 log \frac{P_{on}}{P_{off}} - \alpha_l L$$

where $P_{on}$ and $P_{off}$ are the collected signal powers with and without the pump laser measured at the output fiber respectively, and $\alpha_l$ is the optical propagation loss of $Er^{3+}$-doped LN waveguide in dB/cm. Figs. 4(a) and 4(c) demonstrate the measured spectra with different pomp powers for the signal wavelength of 1530 nm and 1550 nm, respectively.

Figs. 4(b) and 4(d) shows the net gain of the integrated amplifier as a function of the launched pump power for the signal wavelengths at 1530 nm and 1550 nm, respectively. In both cases a rapid rising of gain values following the growing pump power is first observed, which is followed by a slow gain saturation at the higher pump powers (>20 mW). Specifically, the internal net gain for the 1530 nm signal approaches the maximum gain of ~18 dB at the launched pump power of ~40 mW.

The gain dependence of the amplifier on the signal powers was also investigated. By setting the launched pump power to ~20 mW, the net gains provided by the amplifier were measured at different signal powers with the results shown in Fig. 5. It can be clearly seen that the small-signal gain was gradually decreased at increasing signal powers due to the depletion of excited-state population of erbium ions.

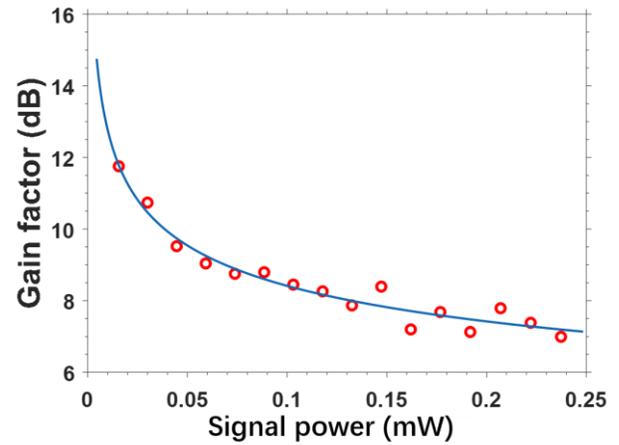

Fig. 5. Gain as a function of launched signal power for wavelength **λ**=1530nm, the launched pump power is about 20 mW.

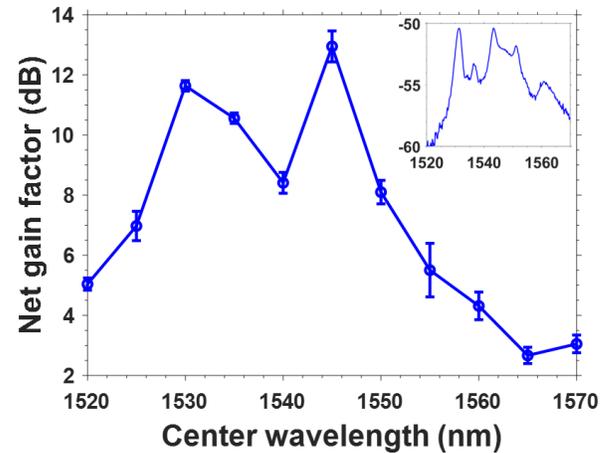

Fig. 6 Net gain as a function of wavelength for a launched pump power of 20 mW. A bidirectional pumping scheme is applied. The reported launched power corresponds to the sum of the powers launched in each of the two input ports. Inset: Amplified spontaneous emission spectrum measured in $Er^{3+}$-doped LN waveguide.

To characterize the gain bandwidth of our waveguide amplifier, the internal net gain measurements were repeated by sweeping the signal wavelength from 1520 nm to 1570 nm at the fixed pump power of 20 mW. The obtained results were shown in Fig. 6. Efficient gains were revealed for the full investigated range of the signal wavelengths, featuring a 3dB gain bandwidth of more than 40 nm centered around 1540 nm. The spectrum of the forward ASE collected at the output fiber without injection of the signal laser was shown in the inset of Fig. 6 for comparison. The good agreement between the gain spectrum and the ASE spectrum corroborates the reliable population inversion of erbium ions excited within the $Er^{3+}$-doped LN waveguide amplifier.

## 3. CONCLUSION

Erbium-doped Ti-diffused lithium niobate channel waveguides have been previously employed for integrated amplifiers and lasers, allowing versatile functionalities due to the excellent electrooptic, acoustooptic, and nonlinear optical properties of the LN substrate [21-23]. However, the large mode size and inhomogeneous distribution of erbium ions in the $Er^{3+}$:Ti:$LiNbO_3$ waveguides limit the attainable gain coefficients to ~2 dB/cm. In contrary, the $Er^{3+}$-doped LN ridge waveguides realized in the current work enable a tight confinement of the guided modes as well as a uniform concentration across the waveguide, facilitating a good spatial overlap between the pump and signal fields with the erbium gain volume and achieving the record-high differential gain of ~5 dB/cm with ~40 mW in-coupled pump power at ~980 nm. Further optimizations concerning the erbium doping concentration to suppress the quenching of excited ions by energy transfer and diffusion processes, the waveguide geometric design to allow for high-power amplifications, and the pumping condition to promote excited-state population, are anticipated to increase the waveguide gain to even higher values comparable with other erbium-doped waveguide amplifiers, holding broad perspectives in sophisticated designs of photonic integrated circuits.

**Funding.** National Key R&D Program of China (2019YFA0705000), National Natural Science Foundation of China (Grant Nos. 12004116, 11874154, 11734009, 11933005, 11874060, 61991444), Shanghai Municipal Science and Technology Major Project (Grant No.2019SHZDZX01).

**Acknowledgment.** We thank Dr. Yang Liu and Prof. Wenxue Li from East China Normal University for their technical support and helpful discussions.

**Disclosures.** The authors declare no conflicts of interest.